\documentclass[
reprint,
aps,
pra,
superscriptaddress,
showpacs, 
nofootinbib,
amsmath,amssymb
]{revtex4-1}

\usepackage{graphicx}
\usepackage{placeins}
\usepackage{epstopdf}

\usepackage{dsfont}
\usepackage{color}
\usepackage[english]{babel}
\usepackage[utf8]{inputenc}
\usepackage[T1]{fontenc}
\usepackage{enumitem}

\begin{document}

\title{Spectral properties and breathing dynamics 
of a few-body Bose-Bose mixture in a 1D harmonic trap}

\author{Maxim Pyzh}
	\email{mpyzh@physnet.uni-hamburg.de}
	\affiliation{Zentrum f\"ur Optische Quantentechnologien, Universit\"at
Hamburg, Luruper Chaussee 149, 22761 Hamburg, Germany}
\author{Sven Kr\"onke}
	\affiliation{Zentrum f\"ur Optische Quantentechnologien, Universit\"at
Hamburg, Luruper Chaussee 149, 22761 Hamburg, Germany}
	\affiliation{The Hamburg Centre for Ultrafast Imaging, Universit\"at 
Hamburg, Luruper Chaussee 149, 22761 Hamburg, Germany}
\author{Christof Weitenberg}
	\affiliation{Institut f\"ur Laserphysik, Universit\"at Hamburg, 
	22761 Hamburg, Germany}
\author{Peter Schmelcher}
	\email{pschmelc@physnet.uni-hamburg.de}
	\affiliation{Zentrum f\"ur Optische Quantentechnologien, Universit\"at 
Hamburg, Luruper Chaussee 149, 22761 Hamburg, Germany}
	\affiliation{The Hamburg Centre for Ultrafast Imaging, Universit\"at 
Hamburg, Luruper Chaussee 149, 22761 Hamburg, Germany}

\date{\today}

\begin{abstract}
We investigate a few-body mixture of two bosonic components, 
each consisting of two particles 
confined in a quasi one-dimensional harmonic trap. 
By means of exact diagonalization
with a correlated basis approach we
obtain the low-energy spectrum and eigenstates
for the whole range of repulsive intra- and inter-component interaction strengths. 
We analyse the eigenvalues 
as a function of the inter-component coupling, covering hereby
all the limiting regimes, and characterize the behaviour 
in-between these regimes by exploiting the symmetries of the Hamiltonian.
Provided with this knowledge we study the breathing dynamics in the
linear-response regime
by slightly quenching the trap frequency
symmetrically for both components.
Depending on the choice of interactions strengths, we identify 1 to 3
monopole modes besides the breathing mode of the center of mass
coordinate. For the uncoupled mixture each 
monopole mode corresponds to the breathing oscillation 
of a specific relative coordinate. Increasing the inter-component
coupling first leads to multi-mode oscillations in each relative coordinate,
which turn into single-mode oscillations of the same frequency 
in the composite-fermionization regime.
\end{abstract}

\maketitle


\section{Introduction}

The physics of ultra-cold atoms has gained a great boost of interest 
since the first experimental realization of 
an atomic Bose-Einstein condensate \cite{Cornell1995,Ketterle1995},
where research topics such as collective modes 
\cite{Cornell1996,Ketterle1996,Stringari1996}, binary 
mixtures \cite{Myatt1997,Cornell1998} and lower-dimensional geometries 
\cite{Olshanii1998,Goerlitz2001,Moritz2003} were in the focus
right from the start. In most of the experiments 
on ultra-cold gases the atoms are but weakly 
correlated and well described by a mean-field (MF) model,
the well-known Gross-Pitaevskii equation (GPE), 
or in case of mixtures by coupled GPEs \cite{Ho1996,Esry1997,Pu1998}. 
Bose-Bose mixtures exhibit richer physics compared to their single component counterpart. 
For instance, different ground state 
profiles can be identified depending on the ratios 
between the intra- and inter-species interaction strengths, 
being experimentally tunable
by e.g.\ Feshbach resonances (FR) \cite{feshbach2010}: 
the miscible (\textit{M}), immiscible symmetry-broken (\textit{SB})
or immiscible core-shell structure, also called phase separation (\textit{PS})
\cite{Papp2008,Hirano2010,McCarron2011}.
Comparing the experimentally obtained densities
to numerical MF calculations \cite{Proukakis2013,Proukakis2016} provides
a sensitive probe for precision measurements
of the scattering lengths
or, if known, the magnetic fields used to tune them \cite{Hirano2010}.
Another possibility to access the interaction regime and thus the scattering lengths
is by exciting the system and extracting the frequencies of low-lying excitations
\cite{Egorov2013}. 
In contrast to a single-species case the collective modes of 
mixtures exhibit new exciting phenomena: 
doublet splitting of the spectrum 
containing in-phase and out-of-phase oscillations, 
mode-softening for increasing inter-component coupling, 
onset of instability of the lowest
dipole mode leading to the \textit{SB} phase 
as well as minima in the breathing mode frequencies w.r.t.\ interaction strength
\cite{Pu1998b,Gordon1998,Morise2000}.

The breathing or monopole mode, 
characterized by expansion and contraction of the atomic density, 
has in particular proven to be a useful tool 
for the diagnostics of static and dynamical properties of physical systems.
It is sensitive to the system's dimensionality, 
spin statistics as well as form and strength of interactions
\cite{Bonitz2009,Bonitz2010,Bonitz2012,Bonitz2014}.
In the early theoretical investigations 
on quasi-one-dimensional single-component condensates 
\cite{Stringari2002}
it was shown that different interaction regimes 
can be distinguished by the breathing mode
frequency, which has been used in experiments
\cite{Moritz2003,Haller2009,Bouchoule2014}. Furthermore, the monopole mode
provides indirect information on the ground state \cite{Mcdonald2013}, 
its compressibility \cite{Altmeyer2007} and the low-lying energy spectrum
such that an analogy has been drawn 
to absorption/emission spectroscopy in molecular
physics \cite{Bonitz2014}.

From a theoretical side, those of the above experiments which are concerned with
quasi-1D set-up are in particular interesting, since correlations are generically
stronger, rendering MF theories often inapplicable. 
Here, confinement induced resonances (CIR)
\cite{Olshanii1998} can be employed to realize the Tonks-Girardeau limit 
\cite{Kinoshita2004,Paredes2004}, 
where the bosons resemble a system of non-interacting fermions in many aspects.
While this case can be solved analytically \cite{Girardeau1960,Girardeau2007},
strong but finite interactions are tractable only to numerical approaches, 
which limits the analysis to few-body systems. 
For instance, a profound investigation of the ground state phases of a few-body
Bose-Bose mixture \cite{Zoellner2008,Hao2009} 
showed striking differences to the MF calculations: for coinciding trap centres,
a new phase with bimodal symmetric density structure,
called composite fermionization (\textit{CF}), is observed while \textit{SB}
is absent for any finite inter-component coupling.
Only in the limit of infinite coupling the ground state becomes two-fold degenerate
enabling to choose between \textit{CF} and \textit{SB} representations
\cite{Zinner2015}, while the MF theory predicts the existence of SB
already for finite couplings. This observation accentuates the necessity to
include correlation effects.  

In this work we solve the time-independent problem 
of the simplest Bose-Bose mixture 
confined in a quasi-1D HO trap with two particles in each 
component, covering the whole 
parameter space of intra- and inter-species interactions, 
thereby complementing the analysis of some previous studies
\cite{Garcia2014,Garcia2014b,Zinner2015}.
To accomplish this, an exact diagonalization method 
based on a correlated basis is introduced.
We unravel how the distinguishability of the components 
renders the spectrum richer and complexer
compared to a single component case \cite{Zoellner2007}.
Furthermore, these results are used to investigate the breathing dynamics of the
composite system. 
While the breathing spectrum of a single component was recently
investigated comprehensively in 
\cite{Schmitz2013,Tschischik2013,Chen2015,Olshanii2015,Zvonarev2015,Bouchoule2016},
reporting a transition from a two mode beating of the center of mass $\Omega_{CM}$ 
and relative motion $\Omega_{rel}$ frequencies for few atoms to a
single mode breathing for many particles, the breathing mode properties of few-body
Bose-Bose mixtures are not characterized so far. For this reason, we 
analyse the number of breathing frequencies and the kind of motion to
which they correspond in dependence on the intra- and inter-component interaction
for the binary mixture at hand.

This work is structured as follows. 
In Sec.\ II we introduce the Hamiltonian of the
system. 
In Sec.\ III we perform a coordinate transformation 
to construct a fast converging
correlated basis. 
Using exact diagonalization 
with respect to this basis 
we study in Sec.\ IV the low-lying
energy spectrum for various interaction regimes.
Sec.\ V is dedicated 
to the breathing dynamics within the linear response regime. 
An experimental realization is discussed in Sec.\ VI and
we conclude the paper with a summary
and an outlook in Sec.\ VII.


\section{Model}
\label{sec:model}

We consider a Bose-Bose mixture containing two components, which are labelled by 
$\sigma \in \{A,B\}$, confined in a highly anisotropic harmonic trap. 
We assume the low temperature regime, where the inter-particle interactions may be modelled 
via a contact potential,
and strong transversal confinement 
allowing us to integrate out frozen degrees of freedom leading to a quasi-1D model. 
Our focus lies on a mixture of $N_{\sigma}=2$ particles, which have the same mass 
$m_{\sigma} \equiv m$ and trapping frequencies 
$\omega_{\sigma,\perp} \equiv \omega_{\perp}$, $\omega_{\sigma,\parallel} \equiv \omega$
in the transversal, longitudinal direction, respectively. 
This can be realized by choosing different hyperfine states of the same atomic species.
By further rescaling the energy and length in units of $\hbar \omega$ and 
$a_{ho}=\sqrt{ \hbar / (m \omega) }$ one arrives at the simplified \mbox{Hamiltonian}:
\begin{equation}
 H = \sum_{\sigma} H_{\sigma} + H_{AB},
\label{eq:tot_hamilt}
\end{equation}
with single-component Hamiltonians $H_{\sigma}$ and inter-component coupling $H_{AB}$.
\begin{eqnarray}
&& H_{\sigma} = \sum_{i=1}^2 
 			\left(
 			-\frac{1}{2} \frac{\partial^2}{\partial x_{\sigma,i}^2}
 			+ \frac{1}{2} x_{\sigma,i}^2
 			\right)
 			\label{eq:species_hamilt}
 			+ g_{\sigma} \delta(x_{\sigma,1}-x_{\sigma,2}) , \\
&& H_{AB} =  g_{AB} \sum_{i,j=1}^2 \delta(x_{A,i}-x_{B,j}) ,
			\label{eq:coupl_hamilt}
\end{eqnarray}
where $g_{\alpha} \approx (2 a_{\alpha}^{3D} \omega_{\perp})/ (\omega a_{ho}) $ with 
$a_{\alpha}^{3D}$ the 3D s-wave scattering length and $\alpha \in \{A,B,AB\}$
are effective (off-resonant) interaction strengths.


\section{Methodology: Exact diagonalization in a correlated basis}
\label{sec:method} 

To obtain information on the low-energy excitation spectrum we employ the 
well-established method of exact diagonalization. 
However, concerning the choice of basis,
instead of taking bosonic number states w.r.t.\ HO eigenstates as in e.g.\ \cite{Garcia2014},
we pursue a different approach by using a correlated atom-pair basis. 
Thereby we can efficiently cover
regimes of very strong intra- and inter-component interaction strengths $g_{\alpha}$
and achieve convergence for relatively small basis sizes of about 700. 

Actually, 
the idea of choosing optimized basis sets to speed up the convergence with respect 
to the size of basis functions can be also seen in the context of the
potential-optimized discrete variable representation (PO-DVR) \cite{po_dvr_1992}.
Here, one employs eigenstates of conveniently constructed
one-dimensional reference Hamiltonians in order to incorporate more
information on the actual Hamiltonian into the basis compared 
to the standard DVR technique \cite{dvr_2000,MCTDH_2000}.
Another approach, stemming from nuclear physics,
uses an effective two-body interaction potential instead of an optimized basis for solving
ultra-cold many-body problems \cite{Reimann2009,Rotureau2013,Zinner2014}.

In order to construct a tailored basis, which already incorporates
intra-component correlations,
let us neglect for a moment the inter-component coupling $H_{AB}$,
which leaves us 
with two independent bosonic components, each consisting of two particles. 
The problem of two particles in a harmonic trap can be solved analytically 
\cite{Busch1998} and boils down to a coordinate transformation and solving 
a Weber differential equation\footnote{ $f''(r)+(\mu+\frac{1}{2}-\frac{1}{4}r^2) f(r)=0$ with $r \in \mathbb{R}$ and $\mu \in \mathbb{R}$} 
under a delta potential constraint. 
Each eigenstate of this bosonic two-particle problem turns out to be a 
tensor product of a HO eigenfunction 
of the center of mass (CM) coordinate 
and a normalized as well as symmetrized Parabolic Cylinder Function\footnote{$D_{\mu}(|r|)=2^{\frac{\mu}{2}} 
e^{-\frac{r^2}{4}} U(-\frac{1}{2} \mu,\frac{1}{2},\frac{1}{2}r^2) $
with $U(a,b,x)$ the Tricomi's hypergeometric function} 
(PCF) $\varphi_n(r)\propto D_{\mu(g,n)}(|r|)$ 
of the relative coordinate 
with $\mu(g,n)$ being a real valued quantum number depending on the interaction strength 
$g$ and excitation level $n \in \mathbb{N}_0$, which is obtained by solving a 
transcendental equation derived from the delta-type constraint:
\begin{equation}
	g=-2^{\frac{3}{2}} \frac
	{\Gamma(\frac{1-\mu}{2})}
	{\Gamma(-\frac{\mu}{2})}.
\end{equation}
Coming back to the binary mixture problem we apply a coordinate transformation 
to the relative frame $\vec{Y}\equiv(R_{CM},R_{AB},r_A,r_B)^T$ defined by:
\begin{itemize}
\item
total CM coordinate \\
$R_{CM}=1/4 \sum_{\sigma} \sum_{i=1}^2 x_{\sigma,i}$,
\item
relative CM coordinate \\
$R_{AB}=1/2 \sum_{i=1}^2 x_{A,i} - 1/2 \sum_{i=1}^2 x_{B,i}$,
\item
relative coordinate for each $\sigma$ component \\
$r_\sigma=x_{\sigma,1}-x_{\sigma,2}$.
\end{itemize}

The Hamiltonian attains a new structure in this frame. 
Firstly, the total CM is separated $H=H_{R_{CM}}+H_{rem}$ 
and is simply a HO of mass $M=4$.
\begin{equation}
	H_{R_{CM}}=
 			- \frac{1}{8} \frac{\partial^2}{\partial R_{CM}^2}
 			+ 2 R_{CM}^2,
 			\label{eq:rcm_hamilt}
\end{equation}
with the spectrum $E_n^{CM}=n+1/2$ and $n \in \mathbb{N}_0$.
The remainder of the Hamiltonian can be decomposed as $H_{rem}=H_0 + g_{AB} H_1$, 
where $H_0=H_{R_{AB}}+\sum_{\sigma} H_{r_{\sigma}}$ can be solved analytically and
$H_1$ couples the eigenstates of $H_0$. 
$H_{R_{AB}}$ is HO of mass $M=1$ and $H_{r_{\sigma}}$ lead to the 
above mentioned Weber differential equations with delta-type constraint.
\begin{eqnarray}
&& H_{R_{AB}}=
 			- \frac{1}{2} \frac{\partial^2}{\partial R_{AB}^2}
 			+ \frac{1}{2} R_{AB}^2, \\
 			\label{eq:rab_hamilt}
&& H_{r_{\sigma}}=
 			- \frac{\partial^2}{\partial r_{\sigma}^2}
 			+ \frac{1}{4} r_{\sigma}^2
 			+ g_{\sigma} \delta(r_\sigma), \\
 			\label{eq:r_hamilt}
&& H_1=\sum_{i,j=1}^2 
		\delta
		\left(
		R_{AB} + (-1)^i \frac{r_A}{2} + (-1)^j \frac{r_B}{2}
		\right). 
		\label{eq:h1_hamilt}
\end{eqnarray}
Now to diagonalize $H_{rem}$ we choose as basis the eigenvectors of $H_0$, which we 
label as $| k,l,m \rangle$ with $k,l,m \in \mathbb{N}_0$. 
The energy of a corresponding basis function and its spatial representation 
are given by:
\begin{eqnarray}
&& \langle R_{AB} , r_A , r_B | k,l,m \rangle = 
\Phi_k^{AB}(R_{AB}) \varphi_l^A(r_A) \varphi_m^B(r_B),
\label{eq:basis_repres}
\\
&& E_{k,l,m}^{(0)} = k+\mu(g_A,l)+\mu(g_B,m)+\frac{3}{2}, 
\label{eq:H0_energy}
\end{eqnarray}
where $\Phi^{AB}_k$ are HO eigenstates and
$\varphi_i^{\sigma}(r) \propto D_{\mu(g_\sigma,i)}(|r|)$.
All $\varphi_i^{\sigma}(r)$ are of even parity because of the bosonic nature of particles 
of each component.

The main challenge now is the calculation of the matrix elements of $H_1$, which are 
complicated 2D integrals at first sight and need to be tackled numerically. 
In the Appendix we provide a circumvention of this problem via the Schmidt decomposition 
\cite{schmidt_decomposition},
allowing us to replace one 2D integral by multiple 1D integrals, 
which results in faster computation times. In quantum chemistry, the algorithm 
for achieving such a representation is known as POTFIT \cite{Jackle1996}.
We will point out several symmetries to avoid the calculation of redundant terms 
and outline (dis)advantages of the whole method.

To summarize, the coordinate transformation to the chosen relative frame (i) decouples
the CM motion and (ii) naturally guides us to employ the analytically known eigenstates of
$H_0$ as the basis states in order to
incorporate intra-component correlations into our basis.


\section{Stationary Properties}
\label{sec:spectrum} 

\begin{figure*}[t]
	\centering
	\includegraphics[width=1\textwidth,keepaspectratio]{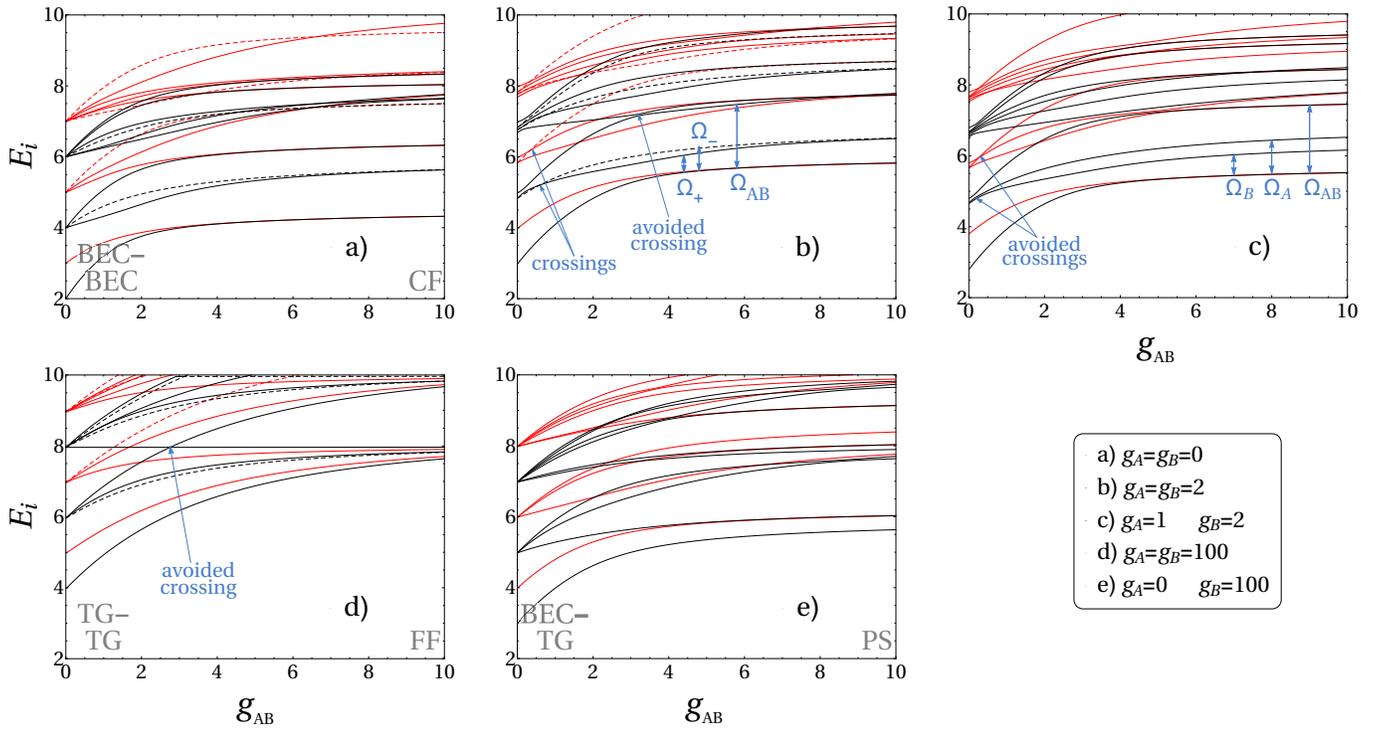}
	\caption{(Colour online) Energy spectrum of $H$ as a function of $g_{AB}$ for different fixed 
	$g_A$ and $g_B$. The decoupled total CM is assumed to be in its ground state. 
	The total parity is thus determined solely by the $R_{AB}$ parity and is marked by 
	black lines (even states) and red lines (odd states). 
	In Fig.\ a) b) d) 
	solid curves correspond to symmetric ($+1$) and dashed to antisymmetric ($-1$) 
	eigenstates under the $S_{r}$ operation.
	The indicated (avoided) crossings are exemplary and simply outline specific features.
	In b) and c), we label the excited states which are relevant for the lowest monopole
	excitations as discussed in Sect.\ \ref{sec:dynamics} by the corresponding excitation
	frequencies $\Omega_{AB}$ etc.
	The labels for the ground-state phases in the limiting regimes 
	\textit{BEC-BEC}, \textit{TG-TG}, etc.\
	follow the nomenclature of \cite{Garcia2014}, see also the main text.
	All quantities are given in HO units.}
	\label{fig:spectrum}
\end{figure*}

By means of the correlated basis introduced above and an efficient strategy for
calculating the Hamiltonian matrix to be diagonalized (see Appendix \ref{sec:app}),
we can easily obtain the static properties of our system for a huge variety
of different intra- and inter-component interaction strengths.
Before going into the details, we need to address the symmetries of $H$ 
in the relative frame.

\subsubsection*{Symmetry analysis}
\label{subsec:sym}

First of all, $H$ commutes with the individual parity operators $P_{Y_i}$ of the 
relative frame coordinates, $[H,P_{Y_i}]=0$.
The eigenvectors of $P_{r_{\sigma}}$ are restricted to even parity because of
the bosonic character of our components. 
Due to the decoupling of $H_{R_{CM}}$ it is sufficient to consider only
the ground state of the total CM motion, which is of even $R_{CM}$ parity, in the following. 
Then, the parity of the $R_{AB}$ degree of freedom completely determines the total parity
of the eigenstates.
Another symmetry arises, if one chooses equal intra-component interaction strengths 
$g_{\sigma}$. 
Under these circumstances the Hamiltonian is invariant under 
$r_A \leftrightarrow r_B$ exchange, which we define as the $S_{r}$ transformation.
It should be noted, that translating all these transformations to the laboratory frame
leads to certain proper and improper rotations of the four-dimensional coordinate space.

\subsubsection*{Energy spectra}
\label{subsec:spectrum}

In Fig.\ \ref{fig:spectrum} we show the total energy spectrum 
as a function of $g_{AB}$ for various fixed values of $g_A$ and $g_B$. 
The total CM is assumed to be in its ground state.
Fig.\ \ref{fig:spectrum}(a) 
depicts the non-interacting intra-component scenario $g_{\sigma}=0$. 
For $g_{AB}=0$ the Hamiltonian represents two
uncoupled non-interacting bosonic species and we will label this regime as 
\textit{BEC-BEC} following the nomenclature of \cite{Garcia2014}. 
The eigenenergies are integers with equal spacings of $\hbar w$, 
which is 1 in our units. 
In this limit the PCFs are even HO eigenstates of mass $m=1/2$. 
The eigenenergies are thus 
$E_{tot}=E_{n=0}^{CM}+E_{k,l,m}^{(0)} \equiv E_{0,k,l,m}^{(0)} = k+2l+2m+2$.

For $g_{AB}=0$, the $i$th eigenenergy corresponding to only even (odd)
$R_{AB}$-parity eigenstates is $(i+2)!/(2!i!)$ fold degenerate with $i \in \mathbb{N}_0$.
Already a small inter-component coupling lifts all these degeneracies such
that branches of eigenenergies arise. In the following, we label these resulting branches
as the $i$th even or odd $R_{AB}$-parity branch, respectively.
Note that this grouping of the energy levels into branches will be used in the following for
all values of $g_{AB}$ and in particular for the analysis of the breathing dynamics in section V.
As we further increase the inter-component coupling strength,
we observe that states corresponding to branches of opposite
$R_{AB}$-parity incidentally cross, 
as they are of different symmetry and consequently not coupled by the 
$H_1$ perturbation.

For very strong $g_{AB}$ values, i.e.\ in the \textit{composite fermionization}
(\textit{CF})
limit \cite{Zoellner2008,Hao2009}, we observe a restoration of degeneracies, but in a different manner, namely
the lowest states merge pairwise forming a two-fold degeneracy. 
In this regime the two components spatially separate for the ground state, 
where one component locates on the left side of the trap, 
while the other is pushed to the right side due to the strong inter-component repulsion.
The two-fold degeneracy of the ground state reflects actually the two possible configurations:
A left B right and A right B left. 
This behaviour can be observed in the relative frame densities, discussed later in this
section. 
Another striking peculiarity for $g_{AB} \rightarrow \infty$
are non-integer eigenvalues and unequal energy spacings. 

A very similar analysis concerning
this specific choice of interactions ($g_{\sigma}=0$ and arbitrary $g_{AB}$)
was performed in \cite{Zinner2015}, 
where an effective interaction approach was employed to greatly improve the convergence 
properties of exact diagonalization in order to access properties of a Bose-Bose mixture up
to $N=10$ particles. However, the analysis only covered a single line of the 
$(g_{A},g_{B},g_{AB})$ parameter space. 
Here, we extend the ground state
analysis of \cite{Garcia2014} and study the low-lying excitations for arbitrary interactions.

In Fig.\ \ref{fig:spectrum}(b) we show the impact of 
moderate but symmetric intra-component interactions of strength $g\equiv g_{\sigma}=2$. 
Already in the uncoupled regime ($g_{AB}=0$) 
we observe fewer degeneracies compared to the $g_{\sigma}=0$ case. 
Nevertheless, we group the
eigenstates into branches of even / odd $R_{AB}$-parity also for finite $g_{\sigma}$
by continuously following the eigenenergies to the $g_{\sigma}\rightarrow0+$ limit.
The reason for the reduced degeneracies is that PCFs are not HO eigenstates any more,
while PCFs of both components are still the same. 
The energy is $E_{0,k,l,m}^{(0)}=k+\mu(g,l)+\mu(g,m)+2$.
To roughly estimate the energetic ordering it is sufficient to know that
the real-valued quantum number $\mu(g,n)$ fulfils for $0<g<\infty$ the following relations:
\begin{itemize}
	\item
	$2n<\mu(g,n)<2n+1$
	\item
	$\mu(g,n)+1<\mu(g,n+1)<\mu(g,n)+2$
\end{itemize}
meaning that a single excitation of the relative motion $r_{\sigma}$
is energetically below a double excitation of the $R_{AB}$ degree of freedom.
E.g.\ the first even $R_{AB}$-parity branch
in the uncoupled 
non-interacting regime (\textit{BEC-BEC} in Fig.\ \ref{fig:spectrum}(a)) 
contains three degenerate states:
$| 2,0,0 \rangle$, $| 0,1,0 \rangle$ and $| 0,0,1 \rangle$ (eq.\ (\ref{eq:basis_repres})). 
By choosing finite $g_{\sigma}$ values $| 2,0,0 \rangle$ acquires a higher energy
than $| 0,1,0 \rangle$ and $| 0,0,1 \rangle$ 
leading to reduced degeneracies in the spectrum.
Another striking feature is the appearance of additional crossings between
states of the same $R_{AB}$-parity due to the $S_{r}$ symmetry. 
States, which possess different quantum numbers concerning the $S_{r}$
transformation ($+1$ or $-1$), are allowed to cross as they 'randomly' do throughout the $g_{AB}$ 
variation. Of course, such crossings are also 
present in the previous non-interacting case, it being also component-symmetric.
An avoided crossing between a state of the first even $R_{AB}$-parity branch and a state of 
the second even $R_{AB}$-parity branch is worth mentioning, 
which is present for all values of $g_{\sigma}$ 
(see the exemplary arrow in Fig.\ \ref{fig:spectrum} (b) or (d)).
States of the same symmetry obviously do not cross according to the 
Wigner-von Neumann non-crossing rule \cite{crossing1929}.

In Fig.\ \ref{fig:spectrum}(c) we asymmetrically
increase the intra-component interactions
$g_{\sigma}$ as compared to the non-interacting case, 
namely to $g_A=1$ and $g_B=2$.
For the uncoupled scenario ($g_{AB}=0$) all the degeneracies are lifted, 
because now PCFs of the $A$ and the $B$ components are different. 
The energy is $E_{0,k,l,m}^{(0)}=k+\mu(g_A,l)+\mu(g_B,m)+2$. 
The energetic state ordering is far from obvious, 
which becomes apparent upon closer inspection of the $\mu(g,n)$ function. 
E.g.\ consider again the first even $R_{AB}$-parity branch.
Its lowest energy state is a single excitation of $r_B$, 
followed by a single excitation of $r_A$. The highest energy of this branch corresponds to a double
$R_{AB}$ excitation. The ordering pattern for higher order branches is even more 
complicated.
For intermediate values of $g_{AB}$ we observe that crossings 
from the previous scenario (with $g_{\sigma}=2$)
between states of the same $R_{AB}$-parity are replaced by avoided crossings
because of the broken $S_{r}$ symmetry. 
The strong coupling regime displays less degeneracies as compared 
to the component-symmetric cases of Fig.\ \ref{fig:spectrum}(a) and (b) with
the two-fold ground state degeneracy remaining untouched.

In Fig.\ \ref{fig:spectrum}(d) we choose very strong intra-component 
interaction strengths $g_{\sigma}=100$.
When the $g_{AB}$-coupling is absent, we have two hard-core bosons in each component. 
The system can thus be mapped to a two-component mixture of non-interacting fermions \cite{Girardeau1960,Girardeau2005} and will be referred to as \textit{TG-TG} limit.
The PCFs become near degenerate with odd HO eigenstates,
which again leads to integer-valued eigenenergies 
$E_{0,k,l,m}^{(0)} \approx k+2l+2m+4$ with equal 
spacings and the same degree of (near-)degeneracies 
as in the non-interacting case
(Fig.\ \ref{fig:spectrum}(a)). 
The limit of strong inter-component coupling displays 
a completely different structure of the spectrum. 
The so-called \textit{full fermionization} (\textit{FF}) 
\cite{Girardeau2007} phase can be mapped to a non-interacting 
ensemble of four fermions with the ground state energy $N^2 /2 = 8$. However, in contrast 
to the single-component case of four bosons, we need to take into account
that the components are distinguishable.
The degeneracy of the ground state is expected to be $N!/(N_A ! N_B !)=6$-fold and 
corresponds to the different possibilities of ordering the laboratory frame coordinates 
while keeping in mind the indistinguishability of particles of each component.
A profound study of these ground states can be done by employing a snippet basis 
\cite{Pfannkuche2008} of $N$ distinguishable particles, which interact with each other 
by an infinite repulsive delta-interaction. 
According to eq.\ (2) of Ref.\ \cite{Pfannkuche2008}, the snippet basis for distinguishable
particles is defined as:
\begin{equation}
	\langle x_1,...,x_N | \Pi \rangle =
	\begin{cases}
       \sqrt{N!} | \Psi_0^F | & \quad \text{if} \quad x_{\Pi(1)} < ... < x_{\Pi(N)}\\
       0            & \quad \text{otherwise,} 
	\end{cases}
\end{equation}
where $\Psi_0^F$ represents 
the ground state of $N$ non-interacting fermions and
$\Pi$ is a permutation of particle coordinates, which defines the sector where $\langle x_1,...,x_N|\Pi\rangle$ has support. 
Every snippet state is one of $N!$ 
possible ground states in the hard-core limit.

To adjust this basis to our case one can follow a procedure described
for a Bose-Fermi mixture in \cite{Minguzzi2011}. The only modification is to replace 
the antisymmetric exchange symmetry of the fermionic species by a symmetric one. 
With respect to the spatial projection
$ \langle \vec{x} | \equiv \langle x_{A,1} , x_{A,2} , x_{B,1} , x_{B,2} |$
we will combine 
the $4!=24$ sectors of the distinguishable case
into $6$ ground states of our Bose-Bose mixture.
Each ground state consists of $4$ different sectors, reflecting the
indistinguishability within each component, meaning that by exchanging the spatial order of
identical particles we switch between these $4$ sectors.
By exchanging the spatial order of distinguishable particles 
we switch between the different ground states.
Thus, the permutations
$\Pi=\tau \ \kappa$ can be decomposed in transpositions $\tau$ 
which exchange identical particles
and transpositions $\kappa$ which exchange distinguishable particles,
leading to the following ground state configurations:
\begin{eqnarray}
	& \Psi_{AABB} &\propto \langle \vec{x} |  (e+(1,2)) (e+(3,4)) \ (e) \rangle, 
	\nonumber \\
	& \Psi_{BBAA} &\propto \langle \vec{x} |  (e+(3,4)) (e+(1,2)) \ (1,3)(2,4) \rangle, 
	\nonumber \\
	& \Psi_{ABAB} &\propto \langle \vec{x} |  (e+(1,3)) (e+(2,4)) \ (2,3) \rangle, 
	\nonumber \\
	& \Psi_{BABA} &\propto \langle \vec{x} |  (e+(2,4)) (e+(1,3)) \ (1,4) \rangle, 
	\nonumber \\   
	& \Psi_{ABBA} &\propto \langle \vec{x} |  (e+(1,4)) (e+(2,3)) \ (2,4) \rangle, 
	\nonumber \\
	& \Psi_{BAAB} &\propto \langle \vec{x} |  (e+(2,3)) (e+(1,4)) \ (1,3) \rangle, 
	\nonumber 
\end{eqnarray}
where the $(i,j)$ notation describes the exchange of particles $i \leftrightarrow j$, $e$ 
is the identity permutation and for two permutations $\Pi_1$, $\Pi_2$,
$\langle \vec{x} |  \Pi_1 + \Pi_2 \rangle$ is defined as 
$\langle \vec{x} |  \Pi_1  \rangle + \langle \vec{x} |  \Pi_2  \rangle$. 
The sub-index of the wave-function $\Psi$ relates to the particle
arrangements in the one-body density distribution.

The last case we discuss is the highly asymmetric case 
$g_A=0$ and $g_B=100$ in Fig.\ \ref{fig:spectrum}(e). 
For $g_{AB}=0$ (\textit{BEC-TG}) one expects, based on the previous considerations, 
integer eigenvalues 
$E_{0,k,l,m}^{(0)} \approx k+2l+2m+3$ and thus
equal spacings as for the cases $g_\sigma=0$ and $g_\sigma=100$ depicted in 
Fig.\ \ref{fig:spectrum}(a) and (d), 
because the PCF of the $A$ component is an even HO eigenstate and the PCF of the $B$ 
component is degenerate with an odd HO eigenstate. 
Very peculiar is the strong coupling case, 
where we observe a non-degenerate ground state, the so-called \textit{phase separation}
(\textit{PS}) phase \cite{Garcia2014}, 
where the $A$ component occupies the centre of the harmonic trap, 
while the $B$ component, in order to reduce its intra-component interaction energy,
forms a shell around the $A$ component.


\subsubsection*{Relative-frame densities}
\label{sec:density}

\begin{figure}[t]
	\centering
	\includegraphics[width=0.48\textwidth,keepaspectratio]{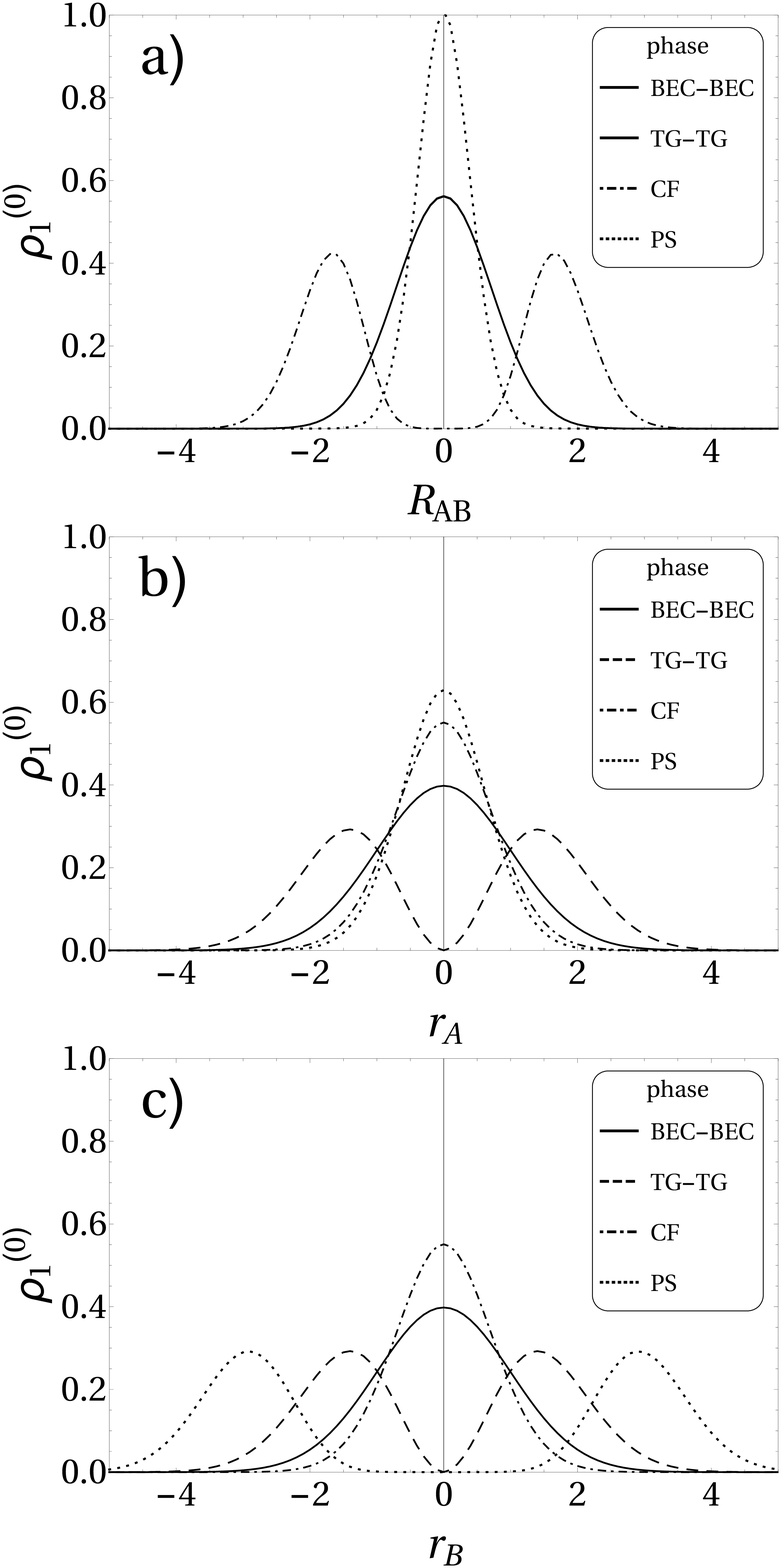}
	\caption{Ground state relative frame probability densities: 
	a) $\rho_1^{(0)}(R_{AB})$ b) $\rho_1^{(0)}(r_A)$ and c) $\rho_1^{(0)}(r_B)$. 
	The depicted limiting
	cases are BEC-BEC ($g_{\sigma}=g_{AB}=0$), TG-TG ($g_{\sigma}=100$, $g_{AB}=0$), 
	CF ($g_{\sigma}=0$, $g_{AB}=10$) and PS ($g_A=0$, $g_B=100$, $g_{AB}=10$). 
	All quantities are given in HO units.}
	\label{fig:density}
\end{figure}

Let us now inspect the relative-frame probability densities
$\rho_{1}(Y_i)$
instead of the usually studied one-body densities $\rho_{1}(x_{\sigma})$ of the laboratory frame
as e.g.\ in \cite{Garcia2014}.
We will see that these quantities can be used to identify
regions of most probable relative distances and provide a more detailed picture of 
particle arrangements than their laboratory frame counterparts.
Moreover, in the quench dynamics study, the subject of the next section, an occupation of a certain 
eigenstate of $H$ will lead to the breathing oscillation of only one relative-frame 
density, making it possible to connect different breathing modes to specific relative
motions within the system, at least for the weakly coupled case $g_{AB} \ll 1$. 

We define these quantities as follows:
\begin{equation}
	\rho_{1}^{(j)}(Y_i)= 
	  \int  \prod_{p \neq i}dY_p \ |\langle \vec{Y} | E_j \rangle|^2,
\end{equation}
where $| E_j \rangle$ is the $j$-th eigenstate of $H$ and we trace out all the
degrees of freedom of the relative frame $\vec{Y}$ except for one. 

Let's compare our results concerning the ground state densities for some limiting cases 
to the ones obtained in \cite{Garcia2014}.
In Fig.\ \ref{fig:density} we show the densities for all the degrees of freedom 
except for $R_{CM}$, which trivially obeys a Gaussian distribution.
In the \textit{BEC-BEC} case all the densities 
are characterized by a Gaussian density profile, 
since the Hamiltonian consists of completely decoupled HOs for each degree of
freedom. The \textit{TG-TG} limit differs from the \textit{BEC-BEC} case
in the $\rho_1^{(0)}({r_\sigma})$ distributions featuring two maxima and a 
minimum in between, a result of strong repulsion within each component. 
This behaviour reflects actually
the already known results of the analytical two-particle solution \cite{Busch1998}, 
where 
$\rho_1^{(0)}(r)$ develops a minimum in the centre of the density distribution for finite $g$
whose value tends to zero as $g \rightarrow \infty$.

The \textit{CF} phase is in some sense a complete counter-part to the TG-TG case. 
Now $\rho_1^{(0)}(R_{AB})$ features two maxima and a minimum in between, 
a result of A and B strongly repelling each other. 
This feature is blurred
in the one-body density distributions $\rho_1^{(0)}(x_{\sigma})$ 
of the laboratory frame and one 
needs to additionally consider the two-body density function 
$\rho_2^{(0)}(x_{A},x_{B})$ to
verify this behaviour \cite{Zoellner2008}.
The density distribution of $\rho_1^{(0)}(r_\sigma)$ is 
more compressed compared to the \textit{BEC-BEC} case due to the 
tighter confinement induced by the other component.

Finally, the \textit{PS} phase corresponds to a core-shell structure, where 
$\rho_1^{(0)}(R_{AB})$ and $\rho_1^{(0)}(r_A)$ show a more pronounced peak, while 
$\rho_1^{(0)}(r_B)$ obeys a bimodal distribution with two density peaks being much further apart 
than both in the \textit{CF} and in the \textit{TG-TG} case.
This can be understood in the following way: 
firstly, the fact that A locates in the trap centre and not the other way around 
is because B needs to minimize its repulsive intra-component interaction
energy by separating its particles.
Secondly, the need to minimize the repulsive inter-component energy 
pushes the B particles even further along the harmonic trap at the cost
of increased potential energy until these two energies balance themselves out.
The two A particles are compressed to closer distances as compared to 
the \textit{BEC-BEC} case because of a tighter trap induced by B, while at the same time
A modifies the HO potential to a double well for B. This results in stronger
localization of particles, which leads to a more pronounced peak in the $R_{AB}$ 
distribution.


\section{Breathing Dynamics}
\label{sec:dynamics}

\begin{figure*}[!]
	\centering
	\includegraphics[width=1 \textwidth,keepaspectratio]{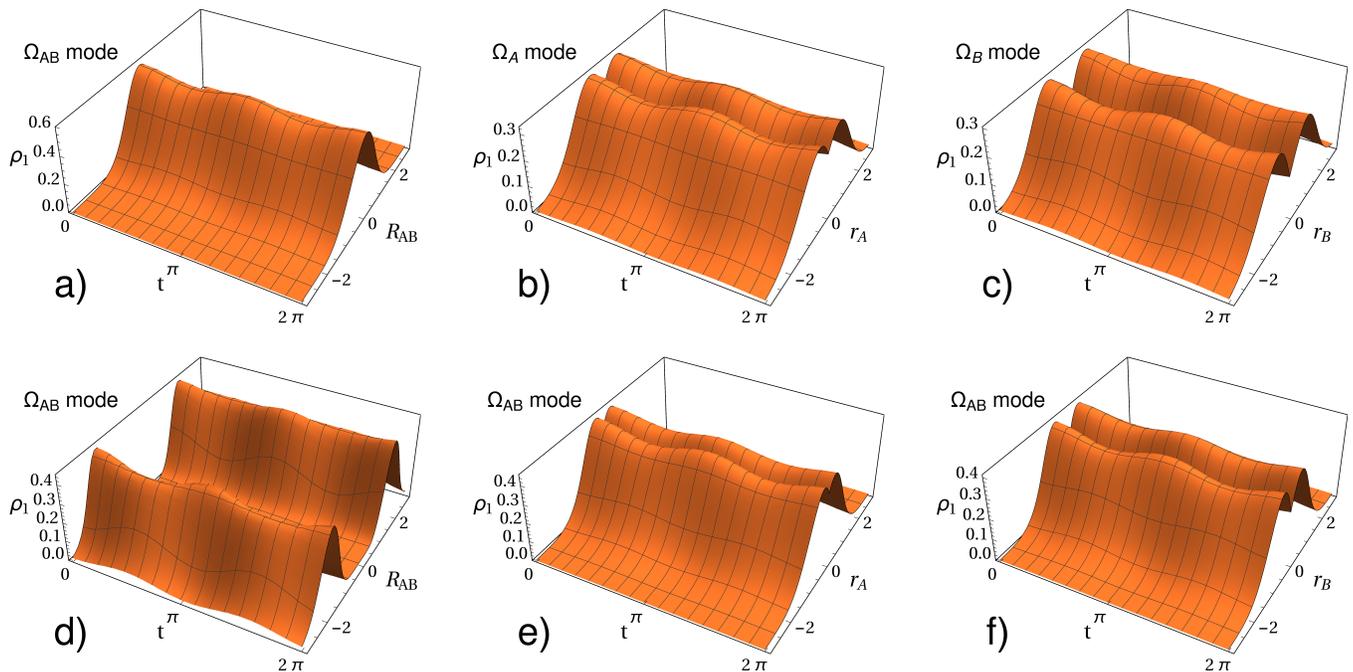}
	\caption{(Colour online)
	Relative frame density modulations
	for the component-asymmetric case of Fig.\ \ref{fig:spectrum} (c). 
	Subfigs.\
	(a)-(c): The decoupled regime $g_{AB}=0$.
	(a) $\rho_1(R_{AB},t)$ oscillates solely due to $|2,0,0 \rangle$ 
	eigenstate with the frequency $\Omega_{AB}=2$,
	(b) $\rho_1(r_A,t)$ due to $|0,1,0 \rangle$ with the frequency $\Omega_{A}$,
	(c) $\rho_1(r_B,t)$ due to $|0,0,1 \rangle$ with the frequency $\Omega_{B}$.  
	Subfigs.\
	(d)-(f): Strongly coupled regime $g_{AB}=10$.
	(d) $\rho_1(R_{AB},t)$,
	(e) $\rho_1(r_A,t)$,
	(f) $\rho_1(r_B,t)$.
	All the profiles oscillate with the same frequency $\Omega_{AB}$. 
	Not shown is the breathing motion of $\rho_1(R_{CM},t)$ with the constant 
	frequency of $\Omega_{CM}=2$,
	it being a decoupled motion of a single-particle HO.
	Quench strength $\delta \omega=-0.1$. 
	All quantities are given in post-quench HO units.	
	}
	\label{fig:density_modu}
\end{figure*}

The spectral properties discussed above can be probed by slightly quenching
a system parameter such that the lowest lying collective modes are excited.
Here, we focus on a slight quench of the trapping frequency in order to
excite the breathing or monopole modes being characterized by a periodic
expansion and compression of the atomic density. While in the single-component
case two lowest lying breathing modes of in general distinct frequencies exist, 
being associated
with a motion of the CM and the relative coordinates \cite{Bonitz2009,Schmitz2013}, 
respectively, the number of 
breathing modes, their frequencies and the associated ``normal coordinates'' 
are so far unknown for the more complex case of a binary few-body mixture and shall be
the subject of this section. 

Experimentally, breathing oscillations
can be studied by measuring the width of $\sigma$ species
density distribution $\int dx_{\sigma} \ x_{\sigma}^2 \rho_1(x_{\sigma},t)$
where we have omitted the subtraction of the mean value 
$\int dx_{\sigma} \ x_{\sigma} \rho_1(x_{\sigma},t)$
squared, which vanishes due to the parity symmetry. From a theoretical point of view,
it is fruitful to define a breathing observable as $\sum_{\sigma,i} x_{\sigma,i}^2$,
whose expectation value is essentially 
the sum of the widths of the A and the B component.

To study the breathing dynamics we will perform a slight and component-symmetric
quench of the HO trapping frequency, where our
HO units will be given 
with respect to the post-quench system.
The initial state for the time-propagation is the ground state 
$|\Psi(t=0)\rangle=|E_0\rangle_{\omega_0}$ of 
some pre-quench Hamiltonian with frequency $\omega_0 \gtrsim 1$.
Then a sudden quench is performed to $\omega=1$. 
The time evolution of this state is thus described
as follows:
\begin{equation}
	|\Psi(t)\rangle = e^{-i H t} |E_0\rangle_{\omega_0} 
	\approx \sum_{j=0}^n c_j e^{-i E_j t}|E_j\rangle,
\label{eq:expansion}
\end{equation}
where $|E_j\rangle \equiv |E_j\rangle_{\omega}$ and
$c_j=\langle E_j|E_0 \rangle_{\omega_0}$ 
is the overlap between the initial state and the $j$th
eigenstate $| E_j \rangle$ of the post-quench Hamiltonian $H$. 
Since both the pre- and post-quench Hamiltonian are time-reversal symmetric,
we assume their eigenstates and thereby also the overlap coefficients $c_j$ to
be real-valued without loss of generality.
A small quench ensures that 
$|c_0| \approx 1$ and only the $n$ lowest excited states are of relevance.
Symmetry considerations further reduce the number of allowed contributions.
E.g.\ states of odd $R_{CM}$-parity or odd $R_{AB}$-parity have zero
overlap with $|E_0 \rangle_{\omega_0}$,
because the initial state is of even $R_{CM}$- and $R_{AB}$-parity and the 
quench does not affect any of the symmetries discussed 
in the previous section.
Similarly, in the component-symmetric case $g_A=g_B$,
states, which are antisymmetric w.r.t.\
the $S_r$ operation,
have no overlap with the pre-quench ground state being symmetric under $S_r$.

For the weakly coupled regime, the relative-frame
coordinates turn out to be extremely helpful for characterizing the participating
breathing modes. Therefore, we study in particular the reduced densities
of the relative-frame coordinates. Employing the 
expansion in post-quench eigenstates from eq.\ (\ref{eq:expansion}), their time-evolution 
may be approximated
within the linear-response regime as
\begin{equation}
	\rho_1(Y_i,t) \approx 
	c_0^2 \rho_1^{(0)}(Y_i)
	+2 \sum_{j=1}^n c_0 c_j  \rho_1^{(0,j)}(Y_i) \cos(\Delta_j t),
	\label{eq:density_osci}
\end{equation}
neglecting terms of the order $c_i c_j$ for $i,j>0$.
So $\rho_1(Y_i,t)$ can be decomposed into
the stationary background 
$\rho_1^{(0)}(Y_i)$ and 
time-dependent modulations of the form  
$\rho_1^{(0,j)}(Y_i)= 
 \int \prod_{p \neq i}dY_p \  \langle\vec{Y}|E_0 \rangle \langle E_j| \vec{Y} \rangle$, 
further called transition densities, 
with oscillation frequency $\Delta_j=E_j-E_0$.

In the following, we regard the excitations of the first even $R_{AB}$-parity 
branch as the lowest monopole modes and show that each monopole mode
is directly connected to the breathing modulation of a single 
relative-frame density, 
if the two components are but weakly coupled.
This behaviour changes for increasing $g_{AB}$, 
where each coordinate begins to exhibit an oscillation
with more than one frequency. By inspecting the modulations 
of the variances of each relative-coordinate
and taking the excitation amplitudes into account, we show that four (three)
breathing modes are excited for $g_A\neq g_B$ ($g_A=g_B$) in the weakly coupled
regime, while only two breathing modes are of relevance in the strongly coupled regime.
First, we inspect
the component-asymmetric case of Fig.\ \ref{fig:spectrum}(c) in detail to illustrate some 
peculiarities of involved breathing modes,
since it contains
the most relevant features. Thereafter,
we unravel differences to the component-symmetric case of Fig.\ \ref{fig:spectrum}(b).

\subsection{Component-asymmetric case}

\begin{figure}[t]
	\centering
	\includegraphics[width=0.45\textwidth,keepaspectratio]{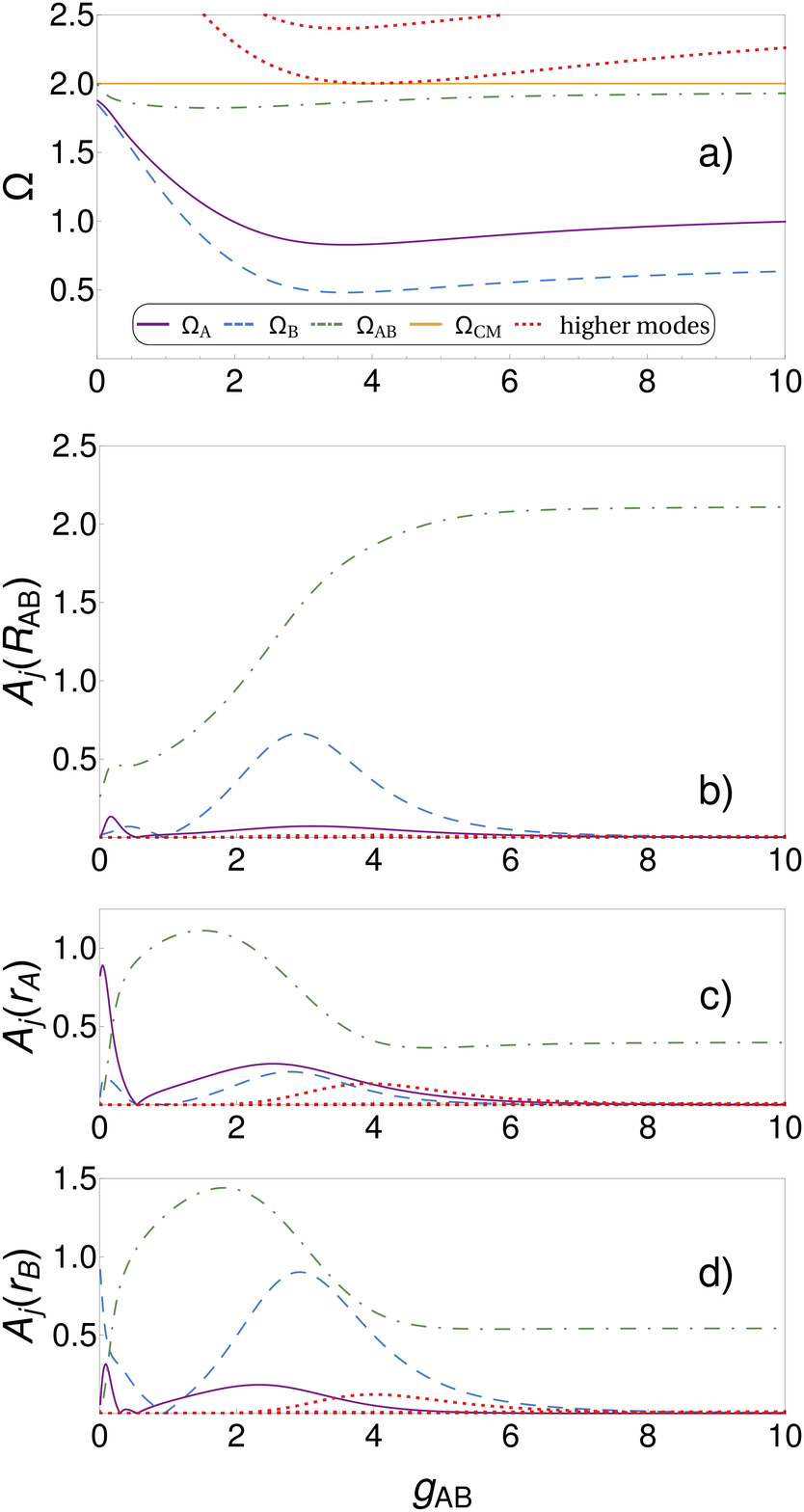}
	\caption{(Colour online)
	(a)
	Breathing mode frequencies as a function of the
	inter-component coupling $g_{AB}$ for the component-asymmetric case  
	$g_A=1.0$ and $g_B=2.0$.
	The modes are labelled with respect to the uncoupled regime, 
	where each mode 
	can be identified with a particular relative-frame motion
	(Fig.\ \ref{fig:density_modu} (a)-(c)).
	(b)-(d)
	Amplitudes $A_j$ (eq.\ (\ref{eq:amplitudes})) of the modes depicted in (a) 
	(same color coding), which determine 
	the relevance of the contribution to the oscillation of the observable $Y_i^2$
	as a function of the inter-component coupling $g_{AB}$: 
	(b) $A_j(R_{AB})$,
	(c) $A_j(r_A)$,
	(d) $A_j(r_B)$.
	The red dotted lines stem from the states of the second even $R_{AB}$-parity branch.
	All quantities are given in post-quench HO units.}
	\label{fig:observables_asym}
\end{figure}

Because of the low amplitude quenching protocol, 
we will excite four breathing modes simultaneously in the
component-asymmetric case ($g_A=1$, $g_B=2$).
Three of them stem from the first even $R_{AB}$-parity branch 
of Fig.\ \ref{fig:spectrum}(c).
Remember, however, that the total CM was assumed to be in the ground state
to keep the spectrum discernible.
One obtains the full spectral picture by including all CM excitations,
meaning duplicating and up-shifting depicted energy curves
by $\Delta E=n$ with $n \in \mathbb{N}$.
This reveals a forth mode, namely a double total CM excitation.
It features the same parity symmetries
and is energetically of the same order as the states 
from the first even $R_{AB}$-parity branch ensuring a considerable overlap
with the initial state. 
The total CM trivially oscillates with the constant frequency 
$\Omega_{CM}=2$ independent
of any interactions $g_{\alpha}$ 
it being a decoupled degree of freedom with the
single-particle HO Hamiltonian (eq.\ (\ref{eq:rcm_hamilt})) \cite{Bonitz2010,Schmitz2013}.
The other three modes,
which are excited, are known analytically, when there is no coupling
between the components, and we label the corresponding 
mode frequencies as:
\begin{enumerate}[label=(\roman*)]
\item
$|0\rangle |0,1,0\rangle \ \leftrightarrow 
\ \Omega_{A} (g_{AB}=0) = \mu(g_A,1)-\mu(g_A,0)$, 
\item
$|0\rangle |0,0,1\rangle \ \leftrightarrow 
\ \Omega_{B} (g_{AB}=0) = \mu(g_B,1)-\mu(g_B,0)$, 
\item
$|0\rangle |2,0,0\rangle \ \leftrightarrow 
\ \Omega_{AB} (g_{AB}=0) = 2$,
\item
$|2\rangle |0,0,0\rangle \ \leftrightarrow 
\ \Omega_{CM}=2$, 
\end{enumerate}
where we have prepended the CM eigenstate $|n\rangle$ 
for a complete characterization of the involved states.
States of higher order even $R_{AB}$-parity branches as well as
higher excitations of the CM coordinate
are negligible 
due to small overlaps with the initial state. 

In the uncoupled regime $g_{AB}=0$, one can show analytically that 
each relative-coordinate density oscillates with a single frequency, each
corresponding to exactly one eigenstate of the first even $R_{AB}$-parity branch
(see Fig.\ \ref{fig:density_modu} (a)-(c)).
E.g.\ for $\rho_1(R_{AB},t)$, the only transition density $\rho_1^{(0,j)}(R_{AB})$
which survives taking the partial trace 
is the one corresponding to $|0\rangle|2,0,0\rangle$, 
while the contributions from the remaining excited states vanish. 
This leads to the breathing
motion in the $R_{AB}$ coordinate with a single frequency $\Omega_{AB}$.
Analogously one can show that $|0\rangle|0,1,0\rangle$ 
solely induces density modulation in $\rho_1(r_A)$
with the frequency $\Omega_{A}$,
while $\rho_1(r_B)$ oscillates with $\Omega_{B}$ exclusively
due to $|0\rangle|0,0,1\rangle$. Thereby, the relative-frame coordinates
render ``normal coordinates'' in the uncoupled regime, which is also
a valid picture for extremely weak
couplings.

By introducing a larger coupling between the components 
one observes that each relative frame density, except for $\rho_1(R_{CM},t)$,
begins to oscillate with up to three frequencies simultaneously. 
So all the modes begin to contribute to the density modulation of each relative
coordinate.
However, there are some peculiarities we observe, for the visualization 
of which the densities are not well suited any more.
Instead, we will transform the breathing observable to the relative frame and
consider the expectation values of individual terms it decomposes into:
\begin{equation}
	 \sum_{\sigma,i} x_{\sigma,i}^2= 4 R_{CM}^2 + R_{AB}^2 + 
	 \frac{1}{2} r_A^2 + \frac{1}{2} r_B^2.
	\label{eq:breathing_oper}
\end{equation}
The expectation value of each observable with respect 
to the time-evolved state $| \Psi(t) \rangle$ 
is directly related to the respective 
relative-frame density:
\begin{equation}
	 \langle \Psi(t) | Y_i^2 | \Psi(t) \rangle = \int dY_i \ Y_i^2 \ \rho_1(Y_i,t).
	 \label{eq:breathing_exp}
\end{equation}
Inserting the time-evolution of the relative frame density from eq.\ (\ref{eq:density_osci})
one finds that the observables decompose into a stationary value 
and a time-dependent modulation as well. 
In particular, we are interested in the amplitudes of modulations, 
when the inter-component coupling is varied, since they determine how many 
frequencies are of essential relevance for the considered motion.
The amplitude of the $j$th mode is essentially composed of the overlap $c_j$ 
and of the transition element:
\begin{equation}
	 \langle E_j | Y_i^2 | E_0 \rangle = \int dY_i \ Y_i^2 \ \rho_1^{(0,j)}(Y_i).
	 \label{transition_elements}
\end{equation}
In order to evaluate the overlaps $c_j=\langle E_j| E_0 \rangle_{\omega_0} $ with $j \neq 0$
in terms of only the post-quench Hamiltonian eigenstates, we perform	
a Taylor approximation with respect to the weak quench strength
$\delta \omega=\omega-\omega_0$, namely $|E_0 \rangle_{\omega_0}
\approx |E_0 \rangle_{\omega}-\delta\omega \frac{d}{d\omega}|E_0 \rangle_{\omega}$
evaluated at $\omega=1$, and arrive at
\begin{equation}
 c_j \approx -\langle E_j| \frac{d}{d \omega} |E_0 \rangle \delta \omega.
\end{equation}
Applying the (off-diagonal) Hellmann-Feynman theorem, one obtains:
\begin{equation}
	\langle E_j| \frac{d}{d \omega} |E_0 \rangle
	= \frac{- \omega \langle E_j | \sum_{i,\sigma} x_{i,\sigma}^2 | E_0 \rangle}
	{E_j-E_0}.
\end{equation}
The overlaps are hence connected to the transition elements of each relative-frame
breathing observable (eq.\ (\ref{eq:breathing_oper})), 
weighted with the inverse of the mode frequency,
which leads to a damping of contributions from higher order branches.
This relation enables us to calculate the amplitude $A_j$, with which the $j$-th mode
contributes to the oscillation of the observable $Y_i^2$:
\begin{equation}
	A_j(Y_i)= 
	\left|
	\frac{c_j}{\delta \omega} \langle E_j | Y_i^2 | E_0 \rangle
	 \right|,
	\label{eq:amplitudes}
\end{equation}
which may be interpreted as the susceptibility of the $Y_i^2$ observable
for the excitation of the state $|E_j\rangle$.

In Fig.\ \ref{fig:observables_asym} (a) we show the values of possible breathing
mode frequencies, obtained from the spectrum of Fig.\ \ref{fig:spectrum} (c). 
$\Omega_{CM}=2$ does not depend on any interactions
$g_{\alpha}$. In contrast to this, $\Omega_{AB}$ is degenerate with 
$\Omega_{CM}$ for $g_{AB}=0$, and when increasing $g_{AB}$, 
decreases to a minimum first and then increases
with the tendency to asymptotically reach $\Omega_{CM}$ again. 
This behaviour strongly
resembles the dependence of the relative-coordinate breathing-mode frequency 
in the single-component case  \cite{Schmitz2013}.
$\Omega_{A}$ and $\Omega_{B}$ 
have qualitatively akin curve shapes, varying much stronger with $g_{AB}$. 
In particular, we note that these frequencies reach values
below the frequency of the CM dipole mode being equal to unity. 

The breathing mode frequencies discussed above are labelled according 
to the peculiarity of the uncoupled regime,
where each eigenstate from the first even $R_{AB}$-parity branch 
leads to a breathing motion of
some specific relative-frame coordinate.
Indeed, if we look at the amplitudes $A_j$ 
in Fig.\ \ref{fig:observables_asym} (b)-(d) in the decoupled regime ($g_{AB}=0$),
we recognize that the amplitude for the coordinate $Y_i$ is non-zero 
only for one mode, namely the one with which $\rho_1(Y_i,t)$ oscillates in 
Fig.\ \ref{fig:density_modu} (a)-(c). 
When we increase the inter-component coupling,
the eigenstates cease to be simple
product states in the relative coordinate frame resulting in 
contamination of each density modulation 
with the frequencies from the other modes as well,
which leads to a three-mode oscillation.
Nevertheless, we label the frequencies 
corresponding to the uncoupled case 
and follow the states continuously throughout the
$g_{AB}$ variation.

Another peculiarity worth noting arises 
in the strongly coupled regime: $\Omega_{\sigma}$
oscillations become strongly suppressed 
for all the observables making $\Omega_{AB}$ and $\Omega_{CM}$
the main contributors to the density modulations.
In Fig.\ \ref{fig:density_modu} (d)-(f) we show that all the relative-coordinate
densities oscillate with the same frequency $\Omega_{AB}$.
We highlight that in the CF regime each density peak of the bimodal distribution
$\rho_1(R_{AB},t)$ does not only breath periodically 
but also its maximum height position
performs dipole-mode like oscillations, see Fig.\ \ref{fig:density_modu} (d).

\subsection{Component-symmetric case}

\begin{figure}[t]
	\centering
	\includegraphics[width=0.45\textwidth,keepaspectratio]{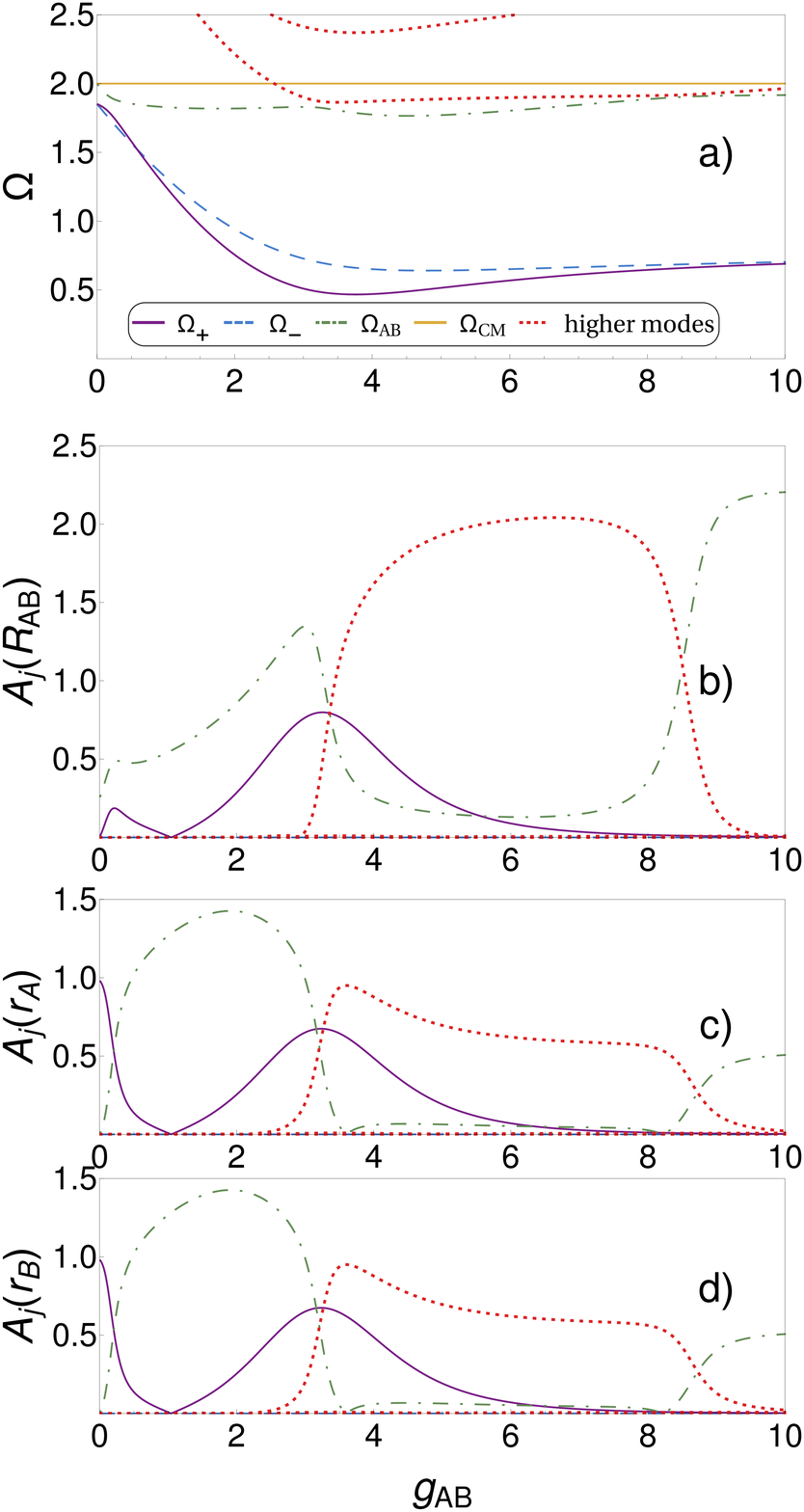}
	\caption{(Colour online)
	(a)
	Breathing mode frequencies as a function of the
	inter-component coupling $g_{AB}$ for the component-symmetric case
	$g_{\sigma}=2.0$.
	(b)-(d)
	Amplitudes $A_j$ (eq.\ (\ref{eq:amplitudes})) of the modes shown in (a) 
	(same color pattern), which determine 
	the relevance of the contribution to the oscillation of the observable $Y_i^2$
	as a function of the inter-component coupling $g_{AB}$: 
	(b) $A_j(R_{AB})$,
	(c) $A_j(r_A)$,
	(d) $A_j(r_B)$.
	The red dotted lines stem from the states of the second even $R_{AB}$-parity branch.
	All quantities are given in post-quench HO units.}
	\label{fig:observables_sym}
\end{figure}

Now we compare the above results with the component-symmetric case of 
Fig.\ \ref{fig:spectrum} (b),
where $g_A=g_B=2$. Fig.\ \ref{fig:observables_sym} (a) 
depicts the possible breathing-mode frequencies.
Here, two main differences arise: 
First, the $\Omega_{AB}$ curve features two minima due to two
avoided crossings with an eigenstate 
of the second even $R_{AB}$-parity branch.
Second, the other two breathing mode frequencies 
of the relative coordinates are degenerate
for $g_{AB}=0$, then separate
with increasing $g_{AB}$ and approach one another asymptotically.
Instead of using the labels $\Omega_A$ and $\Omega_B$ 
as in the component-asymmetric case,
we label these two modes with $\Omega_{+}$ and $\Omega_{-}$, 
since the corresponding excited eigenstates
are symmetric and antisymmetric w.r.t. $S_r$, respectively 
(see Fig.\ \ref{fig:spectrum} (b)).
The latter frequency, however, 
does not give any contribution to the breathing dynamics, 
it being symmetry excluded. 

The motion of the $r_A$ and $r_B$ coordinates is identical due to the imposed
component-symmetry and in the uncoupled regime ($g_{AB}=0$) they both oscillate solely 
with the $\Omega_{+}$ frequency (see Fig.\ \ref{fig:observables_sym} (c) and (d)).
Similarly to the component-asymmetric case, we see  that when increasing $g_{AB}$
the $\Omega_{AB}$ ($\Omega_{+}$) mode contributes also to the observable $r_\sigma^2$ ($R_{AB}^2$).

However, in the intermediate interaction regime ${3.7 \lesssim g_ {AB} \lesssim 8.3}$
we observe a strong suppression
of the $\Omega_{AB}$ mode contribution for all observables, 
which is in stark contrast to the
component-asymmetric case. 
Instead, a state of the second even $R_{AB}$-parity branch gains 
relevance. 
These two eigenstates actually participate in the avoided crossing exemplary 
indicated by an arrow in the spectrum (see Fig.\ \ref{fig:spectrum} (b)) 
and thereby exchange their character. 
By further increasing $g_{AB}$ we observe another exchange of roles, which is attributed
to the presence of a second avoided crossing (see Fig.\ \ref{fig:observables_sym} (a)), 
such that the
strong inter-component coupling regime shows again the absolute dominance of the $\Omega_{AB}$
mode over the other lowest breathing modes besides $\Omega_{CM}$.


\section{Experimental realization}
\label{sec:experiment}

The few-body Bose-Bose mixture studied here should be observable 
with existing cold atom techniques. 
Quantum gas microscopes allow the detection of single particles 
in a well controlled many-body or few-body system \cite{Bakr2010,Sherson2010} 
and recent progress also allows for spin-resolved imaging in 1D systems 
using an expansion in the perpendicular direction \cite{Preiss2015,Boll2016}. 
In these set-ups, the single-particle sensitivity relies 
on pinning the atoms in a deep lattice 
during imaging and experiments have so far focused on lattice systems. 
However, bulk systems might be imaged with high spatial resolution 
by freezing the atomic positions in a lattice before imaging. 
For fast freezing, this would allow a time-resolved measurement of the breathing dynamics. 
Moreover, spin order was recently observed in very small fermionic bulk systems 
via spin-selective spilling to one side of the system \cite{Murmann2015}.

Deterministic preparation of very small samples was demonstrated for fermions 
via trap spilling \cite{Jochim2011} and for bosons by cutting out a subsystem 
of a Mott insulator \cite{Islam2015}. The tight transverse confinement 
for a 1D system can be obtained from a 2D optical lattice, 
while the axial confinement would come from an additional optical potential, 
which can be separately controlled to initialize the breathing mode dynamics.

Choosing two hyperfine states of the same atomic species 
ensures the same mass of the two bosonic species. Possible choices include 
$^{7}Li$, $^{39}K$ or $^{87}Rb$. 
While the former have usable FRs to tune the interaction strengths, 
the latter allows selective tuning via CIR in a spin-dependent transversal confinement
as can be realized for heavier elements. 
Note that the longitudinal confinement needs to be spin-independent in order
to ensure the same longitudinal trap frequencies and trap centres
assumed in the calculations.
The inter-component interaction strength can be tuned via a transverse spatial separation 
as obtained e.g.\ from a magnetic field gradient \cite{Schachenmayer2015}. 
In the case of $^7Li$ and $^{39}K$, 
the inter-component background scattering lengths are negative \cite{feshbach2010} 
leading to negative $g_{AB}$, but although not reported, inter-component FRs might exist. 
Alternatively, $g_{AB}$ might be tuned via a CIR, 
which selectively changes $g_{AB}$ at magnetic fields, 
where the intra-component scattering lengths are very different.

In the following we give concrete numbers for a choice of $^{7}Li$. 
The density distribution has structures on the scale of the HO unit $a_{ho}$ (Fig.\ \ref{fig:density}). 
Choosing the trap parameters as in Ref.\ \cite{Jochim2011} as $\omega / (2 \pi)=1.5$ kHz
and $\omega_{\perp}/(2 \pi)=15$ kHz, yields $a_{ho}=1$ $\mathrm{\mu m}$, i.e.\ 
larger than a typical optical resolution of $0.8$ $\mathrm{\mu m}$. 
At the same time, temperatures much lower than $\hbar \omega/k_B=72$ nK 
are state of the art. 
The breathing dynamics will occur on a time scale of several $100$ $\mathrm{\mu s}$, 
which is easily experimentally accessible.
Choosing a smaller $\omega$ would make the imaging easier, but would impose
stricter requirements on the temperature.

For the observation of the breathing mode dynamics,
one would record the positions of all four particles
in each experimental image and obtain the widths
$\langle R_{CM}^2\rangle_t$, $\langle R_{AB}^2\rangle_t$, $\langle r_{\sigma}^2\rangle_t$
by averaging the occurring relative coordinates
over many single-shots after a fixed hold time $t$.


\section{Discussion and Outlook}
\label{sec:concl}

In this work we have explored
a few-body problem of a Bose-Bose mixture with two atoms 
in each component confined in a quasi-1D HO trapping potential
by exact diagonalization. 
By applying a  
coordinate transformation to a suitable frame 
we have constructed a rapidly converging basis 
consisting of HO and PCF eigenfunctions. The latter stem from the analytical solution
of the relative part of the two-atom problem \cite{Busch1998} 
and include the information about the intra-species
correlations, 
which renders our basis superior to the 
common approach of using HO eigenstates as basis states.

We have then explored the behaviour 
of the low-lying energy spectrum as a function of
the inter-species coupling for various fixed values 
of intra-species interaction strengths.
Hereby we have covered the strongly coupled limiting cases of
Composite Fermionization, Full Fermionization and 
Phase Separation, studied also intermediate symmetric and
asymmetric values of $g_A$ and $g_B$ and related 
the ground state relative-frame densities of some
limiting cases to the known laboratory frame results \cite{Garcia2014}.
We have discussed the evolution of degeneracies and
explained appearing (avoided) crossings
in terms of the symmetries of the Hamiltonian,
which become directly manifest in the chosen relative-coordinate frame. 

Finally, the obtained results were used to study the dynamics of the system
under a slight component-symmetric quench of the trapping potential. 
We have derived expressions
for the time evolution of the relative-frame densities
within the linear response regime and observed that in the uncoupled
regime ($g_{AB}=0$) the density of each relative frame coordinate 
performs breathing oscillations with 
a single frequency corresponding to a specific excited state of the 
first even $R_{AB}$-parity branch of the spectrum. The total CM coordinate
performs breathing oscillations with the frequency $\Omega_{CM}=2$ (HO units). 
For \textit{asymmetric} choices of $g_{\sigma}$ values, three
additional monopole modes participate in the dynamics, 
each of them corresponding to the motion of a particular
relative coordinate:
$\Omega_A$ for the relative coordinate of the A component,
$\Omega_B$ for the relative coordinate of the B component and
$\Omega_{AB}$ for the relative distance of the CMs of both components.
In contrast to this, the \textit{symmetric} case $g_A=g_B$ 
leads to only two additional modes 
because of a symmetry-induced selection rule:
$\Omega_{+}$ for the relative coordinates of both components and
$\Omega_{AB}$ for the relative distance of the CMs of both components.

For not too strong inter-component coupling, each relative coordinate
exhibits multi-mode oscillations and we have explored their relevance 
for the density modulations 
by analysing the behaviour of suitably chosen observables
as one gradually increases the coupling 
between the components for symmetric and
asymmetric choices of intra-component interactions strengths.
Thereby, we have found that for strong couplings, 
where Composite Fermionization 
takes place, the $\Omega_{\sigma}$ ($\Omega_{+}$) modes become highly suppressed, 
leaving only
two monopole modes in this regime: $\Omega_{AB}$ and $\Omega_{CM}$. 
We have observed the same effect for the case of 
Phase Separation (results not shown).
Interestingly, the dependence of 
$\Omega_{AB}$ on $g_{AB}$ strongly resembles the 
behaviour of the relative-coordinate breathing frequency 
in the single-component case
\cite{Schmitz2013}.
All in all, we have obtained 2 to 4 monopole modes 
for the quench dynamics depending 
on the strength of the inter-component coupling and the symmetry of the
intra-species interactions, which is in strong contrast to
the single-component case \cite{Schmitz2013} as well as to the MF results, 
where two 
low-lying breathing modes can be obtained, 
namely an in-phase (out-of-phase) mode for a
component-symmetric (component-asymmetric) quench \cite{Morise2000}.
Finally, we have argued that the experimental 
preparation of the considered few-body mixture
and measurement of the predicted effects are in reach 
by means of state-of-the-art techniques.

This work serves as a useful analysis tool for future few-body experiments.
Measurements of the monopole modes can be mapped to the effective interactions
within the system such that precise measurements of
the scattering lengths or external magnetic fields can be performed. 
The numerical method used here can be applied
to Bose-Fermi and Fermi-Fermi mixtures with 
two particles in each component simplifying
the numerics, because the PCFs have to be replaced by odd HO eigenstates, 
if a bosonic component
is switched to a fermionic one, which significantly 
accelerates the calculation of integrals.
Further, it would be interesting to see how the frequencies and the amplitudes
of the monopole modes vary for an increasing number of particles.
Exploring the spectrum
for negative values of interaction parameters is also a promising 
direction of future research.


\begin{acknowledgments}
The authors acknowledge fruitful
discussions with \mbox{H.-D.} Meyer, J. Schurer, K. Keiler and J. Chen.
C.W. and P.S.
gratefully acknowledge funding
by the Deutsche Forschungsgemeinschaft in the framework of
the SFB 925 ''Light induced dynamics and control of correlated
quantum systems''.
S.K. and P.S. gratefully acknowledge support for this work by the excellence cluster 
''The Hamburg Centre for Ultrafast Imaging-Structure, 
Dynamics and Control of Matter at the Atomic Scale'' of 
the Deutsche Forschungsgemeinschaft.
\end{acknowledgments}


\appendix
\section{}
\label{sec:app}

In the following, we discuss how to efficiently calculate 
matrix elements of the coupling operator $H_1$ 
from eq.\ (\ref{eq:h1_hamilt}) with respect to the basis (\ref{eq:basis_repres}). 
Because of the already mentioned even parity of $\varphi^{\sigma}_i$ one can make simple 
substitutions of the form $\tilde{r}_{\sigma}=-r_{\sigma}$ to show that each delta in the 
sum of $H_1$ gives the same contribution, such that after performing an integral 
over $R_{AB}$ one obtains:
\begin{eqnarray}
	&& \langle a,b,c | H_1 | k,l,m \rangle = \nonumber \\
	&& 4 \int dr_A \int  dr_B \ 
	\Phi_a^{AB} \left( \frac{r_A+r_B}{2} \right)
	\Phi_k^{AB} \left( \frac{r_A+r_B}{2} \right) \nonumber \\
	&& \varphi_b^A(r_A) \varphi_l^A(r_A) 
	\varphi_c^B(r_B) \varphi_m^B(r_B).
\label{eq:perturbation}	
\end{eqnarray}
At this point it is important to notice 
that the integral vanishes for odd $(a+k)$ 
because of the parity symmetries of $\Phi^{AB}$ 
and $\varphi^{\sigma}$, 
which can be seen by transforming 
to relative and center-of-mass coordinates
$r=r_A-r_B$, $R=(r_A+r_B)/2$. 
In the following, 
we assume that the quantum numbers for the $R_{AB}$ coordinate
are restricted to $a,k\in\{1,...,n_{AB}\}$ 
and for the $r_\sigma$ 
coordinates to
$b,c,l,m\in\{1,...,n_{\rm rel}\}$. 
Now our computation strategy consists of three steps:

First, we circumvent evaluating the 
2D integral from eq.\ (\ref{eq:perturbation}) by viewing the
product of the two HO eigenstates $\Phi_a^{AB} \Phi_k^{AB}$ as a pure, 
in general not normalized state $\chi_{a,k}$
depending on the two coordinates $r_A$ and $r_B$
and applying the Schmidt decomposition \cite{schmidt_decomposition} 
or, equivalently, the
so-called POTFIT algorithm \cite{Jackle1996}. 
\begin{eqnarray}\label{eq:potfit}
	\chi_{a,k}(r_A,r_B)
	&& \equiv \Phi_a^{AB} \left( \frac{r_A+r_B}{2} \right)
	\Phi_k^{AB} \left( \frac{r_A+r_B}{2} \right)  \\\nonumber
	&& \approx \sum_{i=0}^{d} \lambda_i^{(a,k)} 
	w_i^{A;(a,k)} (r_A) w_i^{B;(a,k)} (r_B).
\end{eqnarray}
Here\footnote{Note that in contrast to the usual convention 
we do not require the coefficients $\lambda_i^{(a,k)}$
to be semi-positive without loss of generality.}, 
$|\lambda^{(a,k)}_i|^2$  
coincides with the $i$th eigenvalue
of the reduced one-body density matrix corresponding 
to the degree-of-freedom $r_\sigma$
and $w_i^{\sigma;(a,k)} (r_\sigma)$ denotes the corresponding eigenvector, 
which can
be shown to feature a definite parity symmetry. 
We remark that (i) we may choose 
$w_i^{A;(a,k)} (r)= w_i^{B;(a,k)} (r)\equiv w_i^{(a,k)} (r)$
because of $\chi_{a,k}(r_A,r_B)=\chi_{a,k}(r_B,r_A)$ 
without loss of generality and that
(ii) the decomposition of eq.\ (\ref{eq:potfit})
becomes exact for $d\rightarrow\infty$.
Having ordered the coefficients $\lambda^{(a,k)}_i$ 
in decreasing sequence w.r.t. to
their modulus, we choose $d$ such 
that only terms with $|\lambda^{(a,k)}_i / \lambda^{(a,k)}_0| > 10^{-6}$
are taken into account, which results in an accurate approximation to $\chi_{a,k}$.
We perform this decomposition for all the
relevant HO quantum numbers $(a,k)$, meaning
$a \leq k \leq n_{AB}$ with $(a+k)$ even.
This procedure is independent of any interactions 
$g_{\alpha}$ and needs to be executed only once. 

By inserting eq.\ (\ref{eq:potfit}) into eq.\ (\ref{eq:perturbation})
we obtain the following expression:
\begin{eqnarray}
	&& \langle a,b,c | H_1 | k,l,m \rangle \approx \nonumber \\
	&& 4 \sum_{i=0}^d \lambda_i^{(a,k)}
	\int dr_A \ 
	w_i^{(a,k)}(r_A) \varphi_b^{A}(r_A) \varphi_l^{A}(r_A) \nonumber \\
	&& \qquad \qquad \quad \int dr_B \ 
	w_i^{(a,k)}(r_B) \varphi_c^{B}(r_B) \varphi_m^{B}(r_B).
\label{eq:approx}	
\end{eqnarray}
As one can see, the 2D integral is replaced by a sum of products of 1D integrals.
In order to greatly overcome the numerical effort for computing a 2D integral $d$
should be preferably a small number (see below). 

The second step consists in the calculation of 1D integrals and here we provide an efficient strategy
to circumvent redundant computations. Consider the integral:
\begin{equation}
	\int dr \ w_i^{(a,k)} (r) \ \varphi_s^{\sigma} (r) \ \varphi_t^{\sigma} (r) ,
\label{eq:int}	
\end{equation}
with $s \leq t \leq n_{\rm rel}$.
The PCFs are of even parity and thus
only even $w_i^{(a,k)}$ actually contribute 
allowing to reduce the number of expansion terms
in eq.\ (\ref{eq:approx}) to $\lfloor d/2\rfloor$.
Both PCFs $\varphi_s^{\sigma}$ and $\varphi_t^{\sigma}$ 
depend on the same interaction strength 
$g_\sigma$ meaning that the above integral 
does not distinguish between the two subsystems
such that the index $\sigma$ can be dropped for the moment.
Now we fix the PCFs by specifying the strength
of intra-component interaction $g$ and further we fix 
the HO quantum numbers $(a,k)$, which determine the functions $w_i^{(a,k)}$, 
as well as
PCF quantum numbers $(s,t)$. We loop over all $i$ 
and save the integral values, labelled as $(g,a,k,s,t)$. 
This procedure is performed for a set $I$ 
of multiple values of $g$ we are interested in
and for all the relevant quantum number configurations 
$a \leq k \leq n_{AB}$ with $(a+k)$ even and $s \leq t \leq n_{\rm rel}$. 

In the last step we calculate the matrix elements from eq.\ (\ref{eq:approx}).
For HO quantum numbers $(a,k)$ we extract all the expansion coefficients
$\lambda_i^{(a,k)}$, obtained in the first step, and 
for the chosen interactions parameters $(g_A,g_B)$ and 
PCF quantum numbers $(b,c,l,m)$ we pick the appropriate integral values
corresponding to ${(g=g_A,a,k,s=b,t=l)}$ and ${(g=g_B,a,k,s=c,t=m)}$.
The advantage of this procedure is that not only symmetric choices of
intra-component interaction strengths are accessible,
but also an arbitrary asymmetric combination $(g_A,g_B)\in I\times I$.
Additionally,
the proposed scheme can be easily parallelized.
However, adding new $g$ values to the set $I$ 
is in general very time-consuming as one needs to calculate
a bunch of 1D integrals for all the relevant quantum number configurations.

Now let's analyse quantitatively the speed-up obtained by
our algorithm in contrast to the straightforward evaluation of 2D integrals.
Since the energy 
spacing of the PCF modes is approximately 
twice the energy spacing of the HO modes corresponding to the $R_{AB}$ motion,
we assume $n_{\rm rel}=(n_{AB}-1)/2$ with an odd $n_{AB}$ to keep
the number of even and odd $R_{AB}$-parity basis states the same.
The number of 1D integrals one needs to compute for each $g\in I$
in order to construct the $H_1$ matrix 
is approximately $(\overline{d}/64) [(n_{AB}+1)(n_{AB}+3)]^2$,
where $\overline{d}$ is an average number of terms in (\ref{eq:approx}), 
as the criterion $|\lambda^{(a,k)}_i / \lambda^{(a,k)}_0| > 10^{-6}$ requires more terms for larger values of $a,k$.
The number of 2D integrals
amounts to $(1/256)[(n_{AB}+1)(n_{AB}+3)]^3$.
For checking the convergence (see below), we have chosen
$n_{AB}=21$ and $n_{\rm rel}=10$, i.e.\ $2662$ basis states.
The number of expansion terms varies in the interval $d \in \{50,...,100\}$
resulting in $\overline{d}=75$.
Thus, we need to either evaluate $326\ 700$ 1D integrals or $574\ 992$ 2D integrals.
Not only is the number of 1D integrals smaller, 
the computation of one 2D integral
takes also significantly longer than of one 1D integral, 
especially for higher quantum numbers. 
Moreover, in order to build the Hamiltonian matrix for all $(g_A,g_B)\in I\times I$, 
the 2D integrals (\ref{eq:perturbation}) would have to be evaluated 
for the $n_g(n_g+1)/2$ distinct combinations $g_A\leq g_B$, where $n_g$
denotes the cardinality of $I$, while the 1D integrals (\ref{eq:int})
must be calculated only for all $g\in I$, i.e.\ $n_g$ distinct values, which
renders this approach much more efficient.

For the spectra shown in section IV we have chosen $n_{AB}=17$ and $n_{\rm rel}=8$,
i.e.\ $1458$ basis states.
Since we know, that basis states of different $R_{AB}$-symmetry do not couple, we can 
split the $H$ matrix into subspaces of even and odd $R_{AB}$-parity, leading to 
$(729 \times 729)$-size matrices
for each subspace such that
the computational effort for the diagonalization becomes negligible. 
Since the inter-component coupling $g_{AB}$ enters the matrix $H=H_0+g_{AB}H_1$ 
to be diagonalized only
as a pre-factor, a very fine $g_{AB}$ scan can be easily performed. 
We note that for the
covered $g_{AB} \in [0,10]$ space the convergence check 
provides us with the relative energy change below 1\% for the low-lying energy spectrum.


\bibliography{literature}


\end{document}